\documentclass[useAMS,usenatbib]{mn2e}

\usepackage{graphicx}
\usepackage{dcolumn}
\usepackage{bm}
\usepackage{amsmath}
\usepackage{url}

\newcommand{\beq}{\begin{equation}}
\newcommand{\eeq}{\end{equation}}

\def\z22{Z^2_2}

\def\Nbins{{{\cal{N}}_{\rm bins}}}
\def\Npix{{{\cal{N}}_{\rm pix}}}
\def\Pf{P(f)}
\def\Ns{{N_{\rm s}}}
\def\Nbg{{N_{\rm bg}}}
\def\SN{{\cal{S}}}

\def\fb{{f_{\rm b}}}
\def\g{\gamma}
\def\Ftot{{F_{\rm tot}}}
\def\Fb{{F_{\rm bg}}}
\def\Fp{{\Phi_p}}
\def\sp{{\sigma_p}}
\def\O{\Omega}
\def\Aeff{{A_{\rm eff}}}

\def\yvec{{\bf y}}

\begin{document}


\title[Pulsars in the diffuse gamma-ray background]{Extracting the unresolved pulsar contribution to the gamma-ray background}
\author[Alex Geringer-Sameth and Savvas M. Koushiappas]
{
	Alex Geringer-Sameth\thanks{email: alex\_geringer-sameth@brown.edu} and
	Savvas M. Koushiappas\thanks{email: koushiappas@brown.edu} \\
	Department of Physics, Brown University, 182 Hope Street, Providence, RI 02912
}
\date{Submitted 2011 June 23}
\maketitle



\begin{abstract}
We present a statistical framework which can be used to determine the contribution of an unresolved population of pulsars to the gamma-ray background. This formalism is based on the joint analysis of photon time series over extended regions of the sky. We demonstrate the robustness of this technique in controlled simulations of pulsar populations, and show that the Fermi Gamma-ray Space Telescope can be used to detect 
a pulsar contribution as small as 0.1\% of the gamma-ray background. This technique is sensitive to pulsar populations with photon fluxes greater than  
 $\sim 10^{-10}\,\mathrm{cm}^{-2}\,\mathrm{s}^{-1}$. The framework is extensible to arbitrarily complex searches for periodicity and can therefore be tailored to specific applications such as all-sky surveys and   studies of the Galactic center and globular clusters.
\end{abstract}

\begin{keywords}
	methods: statistical --- pulsars: general --- gamma-rays: observations --- surveys --- diffuse radiation --- Galaxy: center
\end{keywords}

\section{Introduction}

The Large Area Telescope (LAT) onboard the Fermi Gamma-ray Space Telescope (FGST or Fermi) is a unique instrument that collects energetic photons from the whole sky, at an energy and spatial resolution as well as in an energy range that offers a new window on high energy astrophysics. In almost three years since the launch of FGST, the sensitivity of the instrument has facilitated the discovery of new classes of objects, including gamma-ray pulsars. Over 80 gamma-ray pulsars have been discovered in the Fermi$-$LAT all-sky data 
(see \cite{2010ApJS..187..460A}, and also \cite{2008Sci...322.1218A,2011arXiv1103.4108A,2009ApJ...699L.102A,2010ApJ...712..957A,2009ApJ...695L..72A,2011ApJ...732...47C,2011ApJS..194...17R,2011ApJ...727L..16R,2010ApJ...725..571S,2010ApJ...714..927A,2010ApJ...712..957A,2011MNRAS.414.1292K}).  

While it is expected that more pulsars will be discovered as the baseline of the experiment is extended, most will remain undetected because their fluxes are below the sensitivity level of current detection techniques.  These pulsars, as a population, contribute to the diffuse gamma-ray background. Untangling the contributions to this background has been a subject of great interest, not only in the context of pulsar physics \citep{2011ApJ...727..123W}, but also in studies aimed at understanding the gamma-ray background near the Galactic center \citep{2011arXiv1106.3583B, 2011PhLB..697..412H,2011JCAP...03..010A,2011arXiv1102.5095D,2010arXiv1012.5839B,2011PhRvD..83h3517H}, as well as a means to extract faint signals from exotic sources, such as dark matter \citep{2010ApJ...722.1939M} and antimatter \citep{2010JCAP...02..016G}.

In this paper we propose a new statistical search strategy that can be used to learn about the {\it cumulative} contribution of pulsars to the gamma-ray background. This technique is an example of a general philosophy/strategy that we advocate, which is based on the concept that even though individual data samples may not contain a detectable source, the statistics of a large number of samples contains information about the sources (see also \cite{2010arXiv1012.1873G}). For the particular case we are studying here, even when a pulsar is not detected within a region of the sky, the data from that region will still contain information. When a large amount of such data is aggregated one can identify a statistical signature of the presence of pulsars even though the individual objects may not pass sensitivity thresholds.

Such a statistical analysis can reveal the properties of the unresolved pulsar population. Application of this technique to  Fermi-LAT data can place bounds on the cumulative contribution of pulsars to the gamma-ray background which are independent of known sources. It is therefore a complimentary approach to the individual studies of bright pulsars with Fermi \citep{1987A&A...175..353B,2006ApJ...652L..49A}.

We begin in Sec.~\ref{sec:methodology} by describing the general strategy that can be used to learn about populations of objects when each individual one is undetectable on its own. We discuss this in the context of the unresolved pulsar contribution to the gamma-ray background. In Sec.~\ref{sec:implementation} we propose a specific implementation involving the statistics of the maximum peaks in a collection of power spectra. It is developed in the framework of classical hypothesis testing, where the goal is to reject the null hypothesis that no pulsars are present in the gamma-ray sky. This includes the development of the statistical tests used to reject this null hypothesis. In Sec.~\ref{sec:fermi} we make predictions for this method as applied to data from  Fermi-LAT and show that under a wide range of circumstances Fermi should be able to discover the presence of unresolved pulsars.  Additionally, we show that individual, flux-unresolved, pulsars may be discovered based only on analysis of their time series. We discuss ways to extract the cumulative pulsar contribution to the background, which requires making assumptions about parameters of describing the pulsar population. Finally, in Sec.~\ref{sec:discussion} we outline how this technique can be generalized to use more powerful tests for periodicity and discuss caveats which can affect the sensitivity of the method.

\section{General Methodology}
\label{sec:methodology}

The detection of a pulsar at high significance relies on statistical tests performed on a collection of photon arrival times. (At radio frequencies, where the vast majority of pulsars have been discovered, the time series is in the radio intensity, not photon counts.) 

For the sake of simplicity assume that a certain statistical test boils down the entire time series into a single number, a "score"\footnote{Throughout this article ``score'' is used in this sense and has nothing to do with the statistical concept of score defined as the derivative of the log-likelihood.}, which is supposed to represent the ``level of periodicity present''. The higher the number the stronger the periodic signal. ``Detecting'' a pulsar is an exercise in classical hypothesis testing and one needs to take into account the fact that even if there is no pulsar present the score may be high because of random chance. Specifically, one needs the probability distribution for the score conditioned on the null hypothesis that there is no pulsar present. The question is asked, ``What are the chances that the score would have been as high as measured if there was no underlying periodicity in the time series?'' If the answer is, for example 0.3\%, then a pulsar is said to be detected at 99.7\% (or ``$3 \sigma$'') significance. In this example, the value of 0.3\% is called the false alarm probability and in practice a $3 \sigma$ detection is hardly convincing. Usually, discoveries are claimed when the false alarm probability is less than $6 \times 10^{-7}$, a ``$5 \sigma$'' detection threshold.

The dominant factor in the detectability of a gamma-ray pulsar is the number of its photons which are collected by the LAT (i.e. the pulsar's photon flux). So far, Fermi has detected pulsars with fluxes as low as $10^{-8}$ cm$^{-2}$ s$^{-1}$ \citep{2010ApJS..187..460A}. These are pulsars whose time series are extremely unlikely to have been generated by a non-periodic process --- unlikely in the sense just discussed. However, it is quite likely that for every pulsar with such a flux the Galaxy contains a great many more with much smaller fluxes. If we assigned a periodicity score to the time series of these faint pulsars the false alarm probabilities would be considerably greater. Most of them would be of order 1. Individually, these pulsations are undetectable with current data and periodicity tests.

However, what if one computes the periodicity score for 40,000 time series, i.e. for every 1 square degree pixel on the sky? A few of these pixels will contain bright pulsars that will be unambiguously detected (these are the pulsars that are discovered using current pulsar search techniques). It is possible that many more pixels  contain pulsars which are not obvious in the data (i.e. their periodicity scores are not  improbably high), while most of the pixels will likely contain no pulsars at all. The goal then is to infer the presence of the undetected population of pulsars. 

The method we propose in this manuscript is based on a very simple observation: The periodicity scores from many separate time series, taken as collection, will be skewed toward larger values due to the presence of pulsars. By analyzing the distribution of scores we can learn about a population of objects whose individual members remain undetected.

This general idea is not limited to the study of the galactic pulsar population. In fact, the concept of analyzing a collection of individually ambiguous signals to learn about a population underlies many studies of diffuse backgrounds. As an example, measuring the empirical counts PDF in sky pixels has been exploited in the study of blazars \citep{2011arXiv1104.0010M,2009PhRvD..80h3504D}, dark matter annihilation in substructure \citep{2011PhRvD..83l3516B,2010PhRvD..82l3511B,2009PhRvD..80h3504D,2009JCAP...07..007L,2008JCAP...10..040S}, as well as pulsars \citep{2010JCAP...01..005F,2010arXiv1011.5501S}. In these cases, the fact that the PDF differs from Poisson indicates that localized sources contribute to the background (even though any single ``hot pixel'' does not constitute a detection of an individual source.) 

A very simple example can illustrate the idea. Imagine we have a collection of 40,000 coins of which 98\% are fair while the other 2\% are rigged to land on heads 90\% of the time. We get to flip each of the coins once and then try to answer the question, ``Are there any unfair coins in this sample?'' On the basis of one flip we have no way of saying whether any individual coin is fair or not. But perhaps the overall distribution of flip results can reveal information about the population of unfair coins. For example, suppose this experiment results in getting the expected number of heads: $40000 \times (0.98 \times 0.5 + 0.02 \times 0.90) = 20320$ heads. We pose the hypothesis test: if the coins were all fair what is the probability of getting 20320 or more heads? The answer is 
\begin{equation}
{\rm P}(\ge 20320) = \sum\limits_{i=20320}^{40000} \binom{40000}{i} \left(0.5\right)^{40000} \simeq 0.0007.
\end{equation}
That is, there is a 0.07\% chance of getting the results we did if every coin were fair. The hypothesis that all the coins are fair has been rejected with greater than $99.9\%$ significance.

Translating this scenario into pulsar language, each coin represents a one square degree patch of the sky. Flipping a coin corresponds to computing the periodicity score from that pixel's photon time series. Heads is a ``high'' score and tails a ``low'' one. If a pixel contains a pulsar the periodicity statistic gives a high score 90\% of the time. The periodicity score for a pixel with no pulsar present has equal chances of being high or low and one can not make any definitive claims based on the results of an individual measurement. However, the cumulative number of ``high periodicity scores'' from all 40,000 square degrees is strongly inconsistent with ``no pulsars''.

\subsection{Cookbook}

The strategy discussed so far is general but can be decomposed into several specific tasks. Here, we will outline the necessary steps, and in Sec.~\ref{sec:fermi} we will develop a specific realization of this procedure which has been designed for application to  Fermi-LAT data.

The first step is to take the gamma-ray events in a region of the sky and divide them into spatially separated time series. This can be done based on a simple pixelization of the sky or by collecting the photon time series from many promising locations (we will address these choices in Sec.~\ref{sec:discussion}). Some preprocessing of the data should also be performed (e.g. applying a barycenter correction to each time series which corrects for the detector's motion with respect to the ``fixed'' solar system barycenter), as well as detector-specific corrections (e.g., see the Fermi Science Support Center\footnote{\tt{http://fermi.gsfc.nasa.gov/ssc/}}).

Next, a periodicity test statistic is chosen and applied to each time series. The choices for the test are numerous. We will detail a straightforward choice in Sec.~\ref{sec:fermi}. In general, the requirement is that one must assign a ``score'' to each time series which in some sense reflects the level of periodicity present. The test should be tailored to the type of objects one is searching for. For millisecond pulsars (MSPs), for example, it may not be necessary to take into account the effects of spin-down (see Sec.~\ref{sec:discussion}).

It is essential to quantify the response of the test statistic to a white noise time series, i.e. an uncorrelated sequence of photons which was not generated by a pulsar. Specifically, one needs the probability distribution for the score under the null hypothesis that no pulsar is present. This is called the null distribution. In the coin flipping example we used above, this probability distribution was ${\rm P}_0(\text{heads}) = {\rm P}_0(\text{tails}) = 0.5$. In some cases the null distribution can be derived analytically. For more complicated periodicity tests the distribution can be found by simply running the periodicity test many times on randomly generated white noise time series.

Finally, given the collection of scores from the various time series, one tests the collection as a whole for deviation from the null distribution. There are a number of statistical tests that can be used for this purpose. Choices include the Kolmogorov-Smirnov and Anderson-Darling tests as well as the traditional $\chi^2$ test of the binned histogram of scores. For the present application, we introduce an additional test, the $A$-test. It is designed to be sensitive to a very small tail of high periodicity scores (see next section and Appendix for more details).

\section{Specific implementation}
\label{sec:implementation}

In this section we present a methodology based on the above strategy. The goal is to detect the presence of unresolved pulsars by jointly examining the photon time series from numerous pixels in some area of the sky. For the sake of simplicity, we will assume that the pulsar period derivatives are very small. This particular implementation is appropriate for a search for the cumulative contribution of MSPs (\cite{2008LRR....11....8L}, \cite{2007AIPC..921...54R}), but can easily be generalized to the case where period derivatives are significant.

\subsection{Choice of periodicity test}

We need a numerical quantity, calculated from the measured photon data from each pixel on the sky, that describes the level of periodicity present in the time series. For this exercise the periodicity score of a time series is chosen to be {\it the normalized peak magnitude of the power spectrum}. We now explain what this quantity represents and how to compute it from a list of discrete photon arrival times.

The Fourier transform is an alternate representation of the time series which highlights the various sinusoidal components that make up the signal. If a pulsar light curve is a pure sine wave its Fourier transform is a delta function spike at the pulse frequency. A well-used technique in pulsar searches is to take the squared magnitude of the complex Fourier transform, called the power spectrum, and search for peaks in this function.
The statistics of the power spectrum for both random data (e.g. \cite{2002AJ....124.1788R}) and for data which contains a signal \cite{1975ApJS...29..285G,1994ApJ...435..362V} have been well studied in general and in the context of pulsar searches.

If photons arrive at times $t_1, t_2, \dots, t_N$ we treat the signal as a train of delta pulses at these times:
\begin{equation*}
s(t) = \sum\limits_{j=1}^{N} \delta(t-t_j).
\end{equation*}
Plugging this into the definition of the continuous-time Fourier transform yields
\begin{equation}
\tilde{s}(f) \equiv \int\limits_{-\infty}^{\infty} e^{-2 \pi i f t} s(t) dt 
		     = \sum\limits_{j=1}^{N} e^{-2 \pi i f t_j}.
\label{eqn:ftdef}
\end{equation}

The unnormalized power spectrum is the absolute square magnitude of the Fourier transform. It is normalized by dividing by the mean power at each value of $f$. For data which contains systematic noise, calculating a running mean is required and may not be trivial. \cite{2002AJ....124.1788R} present several techniques, including using a running mean or a running median (divided by $\ln(2)$ ) to normalize the power spectrum. For gamma-ray data at the high frequencies associated with MSPs there is likely no systematic non-white noise spectrum contaminating the time series. In this case (pure white noise) the mean is simply equal to the number of discrete photon events in the time series. Therefore we search for peaks in the normalized power spectrum $\Pf$ defined as
\begin{eqnarray}
\Pf &\equiv& \frac{1}{N} \left| \tilde{s}(f) \right|^2 \\
		   &=& \frac{1}{N} \left\{ \left[ \sum\limits_{j=1}^{N} \cos(2 \pi i f t_j) \right]^2 + \left[ \sum\limits_{j=1}^{N} \sin(2 \pi i f t_j) \right]^2 \right\}.\nonumber 
\label{eqn:Pdef}
\end{eqnarray}
We are only interested in the maximum of this quantity, and so computationally it is not necessary to store the entire Fourier transform in memory at any one time. This obviates the need for the ~10 billion point Fast Fourier Transforms (FFTs) that would be required for time series that are years long. Instead, one can calculate the power spectrum by making incremental steps in the frequency, only saving the maximum power seen so far. This procedure is trivially parallelized by dividing the frequency interval to be searched into subintervals and searching each of these for its highest peak. \cite{2002AJ....124.1788R} provide trigonometric recurrences which can keep track of the the two sums in Eq. \ref{eqn:Pdef} as $f$ is incremented in small steps without having to compute sines and cosines.

The power spectrum is not an independent quantity for all values of $f$. It is a standard result from the study of discrete Fourier transforms that independent frequency ``bins'' have width $1 / T$, where $T$ is the elapsed time over which the data was taken. For example, a three year LAT observation results in a width of $10^{-8}$ Hz for each independent frequency bin. In searching for MSPs we would like to search over a frequency range corresponding to pulsar periods between, say, 1 ms and 100 ms. In order to perform the search for peaks in the power spectrum we would first compute $\Pf$ starting at $f_{\rm min} = (100\text{ ms})^{-1} = 10 \text{ Hz}$ and then take steps of size\footnote{In practice, one usually searches using a smaller step size in order to accurately explore each potential peak in the power spectrum. However, this does not change the number of {\em independent} frequency bins searched.} $\delta f = 1 / T \simeq 10^{-8} \text{ Hz}$
until reaching $f_{\rm max} = (1\text{ ms})^{-1} = 1000 \text{ Hz}$. Therefore, the exploration of the normalized power spectrum for each time series requires searching $\Nbins$ frequency bins, where 
\begin{equation}
\Nbins = (f_{\rm max} - f_{\rm min} ) T \approx 9 \times 10^{10}.
\label{eqn:Nbins}
\end{equation}

In general, pulsar light curves are more complicated than sine waves which results in the Fourier transform having a series of spikes at integer multiples of the pulsar frequency. This fact motivates many pulsar searches to look for spikes in the  sum of the first $k$ harmonics of the power spectrum. Here we perform a more simple analysis that does not include the statistical details of searching the harmonic sum. However in practice, the pulsar search may be more sensitive if the highest harmonic-summed peak is used as the test statistic. We defer the discussion of various choices for the test statistic to a later section. 

In summary, we compute the normalized power spectrum for the photon arrival time series for each pixel on the sky. The peak power in the power spectrum (in the frequency range of interest) is assigned to that pixel as its ``periodicity score''. We will now explore the probability distributions describing the scores. 

\subsection{Statistics of the power spectrum peak for random data}

For each pixel the maximum of the power spectrum is a random variable. Following standard notation we call the random variable $X$. A specific realization (or measurement) of $X$ is denoted by a lowercase $x$. If a pixel does not contain a pulsar, we assume that its power spectrum is just white noise, i.e. there are no periodic signals present in the frequency range of interest. In this case, the normalized power in each independent frequency bin is distributed according to an exponential distribution with a mean of 1 (e.g.~\cite{2002AJ....124.1788R}).

Under the null hypothesis of no pulsars the score $X$ is the maximum of $\Nbins$ independent exponentially distributed random variables. The cumulative distribution function (CDF) $F(x)$ is the probability that all of the $\Nbins$ random variables are less than $x$. This is simply equal to $[F_1(x)]^\Nbins$, where $F_1(x) = 1-\exp(-x)$ is the CDF for a single exponentially distributed variable. The value of $\Nbins$ is large (Eq.~\ref{eqn:Nbins}) and we can therefore make the following approximation,
\begin{eqnarray}
F(x) &=& \left[1 - \exp(-x) \right]^\Nbins \nonumber \\
     &=& \left[1 - \frac{e^{-(x - \log\Nbins)}}{\Nbins} \right]^\Nbins \nonumber \\
     &\simeq& e^{-e^{-(x - \log\Nbins)}}.
\label{eqn:CDF}
\end{eqnarray}
This result holds to high precision when $\Nbins \sim 10^{10}$. 

The above expression shows that $X$ is distributed according to what is known as a Gumbel distribution, sometimes called an ``extreme value distribution''. The probability distribution falls off extremely rapidly to the left of the mode at $x=\log\Nbins$ and has a less steep tail to the right. Because $\log\Nbins$ is a location parameter of the distribution the width of the Gumbel distribution does not change as $\Nbins$ increases. Also note that as the observation time increases the distribution shifts to the right at a logarithmic rate. This has important consequences that we discuss later. Looking ahead, as the observation time $T$ increases, a pulsar's power will grow in proportion to $T$ while the random power it competes with grows only as $\log T$.

It is easy to invert $F(x)$ to find 
\begin{equation}
x = \log\Nbins -\log(-\log F).
\label{eqn:inverseCDF}
\end{equation}
Therefore, given a uniform deviate $F$ between 0 and 1, Eq.~\ref{eqn:inverseCDF} can be used to transform it into a Gumbel distributed random variable.

\subsection{Statistics of the power spectrum peak when a pulsar is present}

The only distribution needed in order to perform an experiment that tests whether pulsars are present in the gamma-ray background is the null distribution given by Eq. \ref{eqn:CDF}. The test is simply whether the collection of time series is consistent with none of them containing any pulsar signal. In that case the score $X$ for each time series is distributed as Eq. \ref{eqn:CDF}.

However, in order to test the sensitivity of this method we need to be able to simulate situations where pulsars are present in the sky. In fact, to learn anything about the details of the pulsar population one needs some sort of model for the way pulsars contribute to the background. Here we discuss how the presence of a pulsar affects the  chosen periodicity statistic. We will return later to the question of extracting population parameters from the time series data.

When a pulsar contributes photons to the time series, the peak of the power spectrum is distributed differently.  In this case $X$ is distributed as the maximum of two variables. The first is a random variable representing the power in the bin at the pulsar's frequency. The second is a Gumbel distributed variable corresponding to the maximum power in the other $(\Nbins -1)$ frequency bins. For frequency bins which are not at the pulsar's frequency, the pulsar photons contribute to the Fourier transform as if they were randomly distributed along with all the other photons. That is, the normalized power spectrum for the $(\Nbins -1)$ other frequency bins is a white noise spectrum. We have already shown that the maximum power that will be found in these $(\Nbins -1)$ bins is distributed according to $F(x)$ (Eq. \ref{eqn:CDF}).

In order to determine the height of the normalized power spectrum for the bin at the pulsar's frequency we have to go back to the definition of the Fourier transform\footnote{This paragraph is based on the geometric interpretation given in \cite{1994ApJ...435..362V}.}. The Fourier transform (Eq. \ref{eqn:ftdef}) is seen to be the sum of unit vectors in the complex plane, one vector for each photon in the time series. In the case of white noise, each of these $N$ vectors has a random direction and the sum can be thought of as the endpoint of a random walk. This gives rise to the power in one frequency bin being distributed according to the exponential distribution with scale parameter $N$. More precisely, let $\yvec$ be the sum of $N$ randomly directed 2-dimensional unit vectors. The direction of $\yvec$ will be uniformly distributed between $0$ and $2 \pi$. The squared length of $\yvec$ will be distributed according to
\begin{equation}
{\rm Prob}(\zeta < |\yvec|^2 < \zeta + d \zeta) = \frac{e^{- \zeta / N}}{N}  d \zeta
\label{eqn:Prand}
\end{equation}
It is easy to see that the normalized power in such a frequency bin, given by $|\yvec|^2 / N$, is exponentially distributed with scale parameter equal to 1, as stated above.

Consider now a time series where $\Ns$ photons come from a pulsar and $\Nbg$ are uncorrelated background photons, such that the total number of photons is $N=\Ns + \Nbg$. We examine the Fourier bin at the pulsar's frequency and consider the idealized case where all the pulsar power lies in this single frequency bin with no power in harmonics. In this case each vector in the sum in Eq. \ref{eqn:ftdef} over the $\Ns$ pulsar photons points in the same direction. It therefore has a length equal to $\Ns$. The other $\Nbg$ background photons point in random directions and their sum in the Fourier transform is given by a randomly directed vector whose squared length $l$ is distributed according to Eq. \ref{eqn:Prand} with $N$ replaced by $\Nbg$. To get the value of the normalized power spectrum for this frequency bin we take the squared length of the sum of the ``signal vector'' and the ``background photon vector'' and divide by the total number of photons in the time series. Defining $P_p$ to be the normalized power in the frequency bin at the pulsar's frequency we have
\begin{equation*}
P_p = \frac{1}{N} \left[ \Ns^2 + l + 2 \Ns \sqrt{l} \cos(\theta) \right].
\end{equation*}
The power spectrum height is seen to be a random variable: the quantity $l$ is distributed as $l \sim (1/\Nbg) \exp(-l/\Nbg)$ and $\theta$ is a uniform random variable between $0$ and $2 \pi$.

We introduce the following new variables:
\begin{equation}
\SN \equiv \frac{\Ns}{\sqrt{N}} = \frac{\Ns}{\sqrt{\Ns+\Nbg}}        \label{eqn:SNdef} 
\end{equation}
\begin{equation}
\fb \equiv \frac{\Nbg}{N} = \frac{\Nbg}{\Ns+\Nbg}                    \label{eqn:fbdef}
\end{equation}
The first can be thought of as a signal to noise term representing how many photons in a pixel are due to a pulsar vs. background. The second measures the fraction of photons in a pixel which are not due to the pulsar. In terms of these the normalized power spectrum becomes
\begin{equation}
P_p = \alpha \, \SN ^2 + l' + 2 \sqrt{\alpha}\, \SN \, \sqrt{l'} \cos \theta,
\label{eqn:Pp}
\end{equation}
where the variables $\theta$ and $l'$ are distributed according to
\begin{equation}
l'     \sim \frac{1}{\fb} e^{-l' / \fb} \nonumber 
\end{equation}
\begin{equation}
\theta \sim \text{Uniform}[0,2\pi]. \nonumber
\end{equation}

In Eq.~\ref{eqn:Pp}, $\alpha$ is introduced to take into account the fraction of the pulsar's total power that lies in this single frequency bin. If the light curve of the pulsar were a perfect sine wave all of the signal power would lie in the bin at the fundamental frequency and $\alpha=1$. In more realistic situations the power will be divided up into higher harmonics and $\alpha$ may be less than 1.

For reference, we note that the probability distribution of $P_p$ has been worked out analytically in \cite{1975ApJS...29..285G}, which also contains general results that may be of use when considering more complicated tests for periodicity. In particular, the probability distribution for the sum of an arbitrary number of harmonics in the power spectrum is also derived.

\subsection{Rejecting the null hypothesis of ``No pulsars''}

As described above, each sky pixel is assigned a periodicity score $X$ which is defined to be the peak height of its normalized power spectrum. The goal is to take this collection of $X$ values and perform a statistical test of the following null hypothesis: The time series for every pixel is nothing but white noise, i.e. no pulsars are present in any of the pixels. More precisely, we ask if the collection of measured $X$ values is consistent with each score being drawn from the distribution in Eq. \ref{eqn:CDF} (i.e. generated by a random white noise time series).

This task can be accomplished by a number of statistical methods. Here, we use a new test developed specifically for this application. In this section we outline how the test works and refer the reader to the Appendix for details. 

It is desirable (and possible) to use a classical hypothesis test to learn about the sensitivity of this method. The idea is to boil the collection of measured $X$ values into a single test statistic we call $A$. The quantity $A$ should, in some sense, indicate the overall level of periodicity present in the gamma-ray sky, just as $X$ did for a single pixel. Small values of $A$ should indicate ``less evidence for periodicity'' than do large values of $A$.

The ``$A$ test'' is based on the quantity (see Appendix), 
\begin{equation}
A = \frac{1}{\sqrt{N}} \left\{ \left[ \sum\limits_{i=1}^{N} -\log\left[1-F(x_i) \right] \right] - N\log N  + \log N!  \right\},
\label{eqn:Amini}
\end{equation}
where the $x_i$ are the measured scores (normalized power spectrum peaks) for each of the $N$ time series and $F(x)$ is the CDF of the null distribution given by Eq.~\ref{eqn:CDF}. The test is designed to give more weight to time series with large scores.

The test statistic is treated as a random variable and its probability distribution under the null hypothesis (that every sky pixel contained only non-periodic, random photons) is quantified. A significance threshold is chosen and the critical value $A^*$ is defined so that if the null hypothesis holds, then the probability that $A < A^*$ equals the chosen significance. For example, if we want to perform a ``$3 \sigma$'' search one finds $A^*$ such that ${\rm P}(A<A^*)=0.997$. If we find that the observed value of $A$ is in fact greater than $A^*$ the null hypothesis is to be rejected at ``$3 \sigma$'' significance. In other words, it would be extremely unlikely to measure such a high value of $A$ if there were no pulsars. This indicates that pulsars contribute to the gamma-ray background.

\section{Application to \tt{Fermi-LAT}}
\label{sec:fermi}

We now turn to the question of detecting the presence of pulsars in the gamma-ray sky using current data. We assess the conditions where the proposed formalism is successful in rejecting the null hypothesis of ``no pulsars'' in the diffuse background as measured by the LAT instrument on board Fermi. In this section, we will demonstrate the robustness of this method by generating simulations  which contain a controlled population of pulsars with known properties. We utilize the maximum normalized power periodicity test along with the $A$ test as described above.

Assume that a region of the sky is isotropically populated with pulsars that all have the same flux, $\Fp$, defined as photons per area per time in some energy range. These pulsars contribute a fraction $\g$ of all the photons received by the LAT in this energy range. That is, of all the photons that LAT detects over the entire sky a fraction $\g$ of these originated from pulsars each having a flux $\Fp$. The projected number density of pulsars is given by $\sp$ (number of pulsars per solid angle).

The average flux the LAT measures is given by $\Ftot$ in units of photons per area per time per solid angle  (in the relevant energy range). In addition to pulsars we assume a uniform, isotropic background flux $\Fb$ (same units as $\Ftot$). The independent parameters of this model are $\Fp$ and $\g$. The background flux is chosen to make up the difference between the pulsar contribution and the observed total flux. Specifically,
\begin{equation}
\Ftot = \Fb + \sp \Fp,
\label{eqn:Ftotdef}
\end{equation}
and
\begin{equation}
\g \Ftot = \sp \Fp.
\label{eqn:gdef}
\end{equation}
These two equations determine $\Fb$ and $\sp$ in terms of $\Fp$, $\g$, and the observed $\Ftot$. These equations are more easily interpreted by multiplying through by the solid angle of the survey and by the observation time and effective area of the detector. Then $\Ftot$ becomes the total number of photons received by the LAT over the entire survey area, $\sp$ becomes the total number of pulsars in the survey area, $\Fp$ the number of photons received from each pulsar, and $\Fb$ the total number of background (non-pulsar) photons received over the survey area. Solving the above equations we find
\begin{equation}
\Fb = (1 - \g) \Ftot,
\label{eqn:Fbdef}
\end{equation}
and
\begin{equation}
\sp = \frac{\g \Ftot}{\Fp}.
\label{eqn:spdef}
\end{equation}

We assume a value of $\Ftot = 8.72 \times 10^{-10} \mathrm{cm}^{-2} \mathrm{s}^{-1} \mathrm{deg}^{-2}$, in the energy range $[0.8-6.4]$GeV \citep{2010PhRvL.104j1101A}. This includes the energy range in which pulsars are most important relative to the total flux \citep{2010ApJS..187..460A}. 

In order to generate simulated data, we need a survey area and pixel size. We choose the pixel size, $\O$, to be 1 square degree, and we will use two choices for the survey area: 40,000 square degrees which represents the all-sky survey, and 1,000 square degrees, which roughly represents the inner $\sim 32 \times 32$ degrees around a region such as the Galactic center. 

We must evaluate Eqs. \ref{eqn:SNdef} and \ref{eqn:fbdef} to generate a normalized power for each pixel that contains a pulsar. The number of background photons in a pixel is
\begin{equation}
\Nbg = \Fb \, \O \, \Aeff \,  T,
\end{equation}
where $\Aeff$ is the (orbit-averaged) effective area of  LAT (2000 cm$^2$) and $T$ is the observation time (3 years). The number of pulsar (signal) photons in a pixel which contains a pulsar is
\begin{equation}
\Ns = \Fp \, \Aeff \, T.
\end{equation}
Inserting these quantities in Eqs.~\ref{eqn:SNdef} \& \ref{eqn:fbdef}, we have
\begin{eqnarray}
\SN &=& \sqrt{ \frac{\Fp}{\Fb\O + \Fp} \Fp \Aeff T} \\
    &=& \left(1 - \g + \frac{\Fp}{\Ftot \O}\right)^{-1/2} \left( \frac{\Fp}{\Ftot \O} \right) \sqrt{ \Ftot \, \O \, \Aeff \, T},\nonumber 
\label{eqn:SNsim}
\end{eqnarray}
and
\begin{equation}
\fb = \frac{1-\g}{1-\g + (\Fp /\Ftot \O)}.
\label{eqn:fbsim}
\end{equation}

For a given choice of $\Fp$ and $\g$ we can use these last two equations along with Eq.~\ref{eqn:Pp} to generate a normalized power in a pixel that contains a pulsar \footnote{There are many choices for $\Fp$ and $\g$ that give a number of pulsars which is larger than the number of pixels, i.e. $\sp \O > 1$. When this is the case we need to generate a normalized power for each pulsar in the pixel, a peak power from the other $\sim \Nbins$ frequency bins and then take the maximum of all these to be the periodicity score $X$ for the pixel. We have found that for the range of parameter space we discuss the extra pulsars in each pixel do not change the results. Therefore, we run the simulations with at most one pulsar per pixel (though $\sp$ is allowed to be greater than 1).}. For simplicity the simulations were performed using $\alpha = 1$. Consequences of relaxing this assumption will be discussed later.

We explore the parameter space to see when pulsars will be detected by this method using the $A$ statistic defined above. We choose a value of $A^*$ corresponding to a 99.7\% (``$3 \sigma$'') detection. For each pair of values $\Fp$ and $\g$ we create 1,000 realizations. For each realization we simulate 40,000 (all-sky) and 1,000 (Galactic center) values of $X$ (one for each pixel) and compute the $A$ statistic. Out of the 1,000 trials we count the number in which the null hypothesis is rejected. The fraction of trials in which the null hypothesis is rejected is the sensitivity (or power) of the proposed test. For example, if for a particular choice of $\Fp$ and $\g$ we find that in 900 out of 1,000 simulations the null hypothesis is rejected (i.e. $A > A^*$ in 900 of the simulations), then there is an 90\% chance of making a ``$3 \sigma$'' detection of the presence of pulsars.

\subsection{Results}

\begin{figure}
\begin{center}
\includegraphics[height=7.cm]{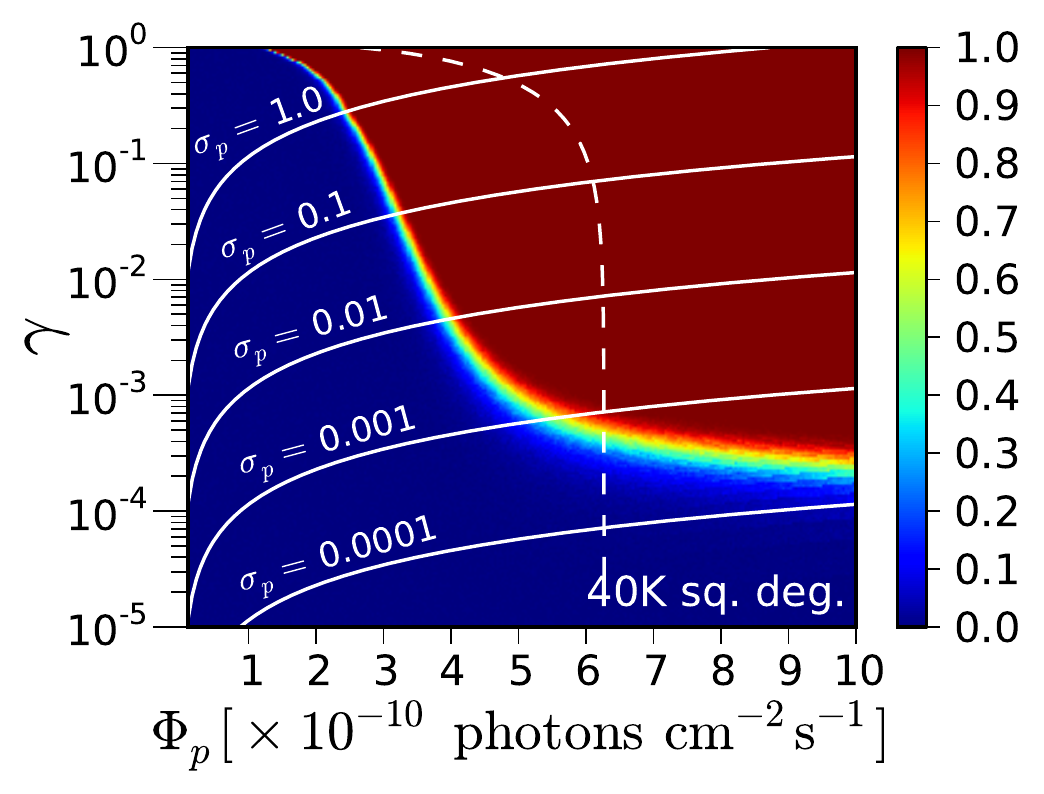}
\caption{A demonstration of the statistical power of the method to detect the presence of pulsars over the entire sky. The color coding represents the probability of rejecting the null hypothesis of ``no pulsars'' at 99.7\% significance. $\Fp$ is the photon flux of each individual pulsar in the energy range $[0.8-6.4] \mathrm{GeV}$. The quantity $\g$ represents the fraction of the total gamma-ray background due to pulsars. Solid contours give the number density of pulsars (in units of pulsars per square degree). The proposed method can reveal the presence of a pulsar population contributing as little as $10^{-3}$ of the diffuse gamma-ray background. Note that, within the range of pulsar fluxes shown, every individual pulsar is {\em flux}-unresolved because $\Fp$ is less than LAT's point source sensitivity threshold. Many of these flux-unresolved sources may be individually discovered based solely on an analysis of their time series: the dashed line represents the $5\sigma$ detection threshold for individual pulsars based on the height of their power spectrum peak (see text for details). 
}
\label{fig:40k}
\end{center}
\end{figure}

Figure \ref{fig:40k} shows the results of the parameter space scan over values of $\Fp$ below $10^{-9}$ cm$^2$ s$^{-1}$ and over the full range of $\g$ from $10^{-5}$ to 1 for a simulated all-sky survey of 40,000 square degrees.  The colour-coding corresponds to the power of this method to reject the null hypothesis that there are no pulsars at 99.7\% (``$3 \sigma$'') significance. In the dark red region the null hypothesis is practically guaranteed to be rejected. In the blue region the null hypothesis will be rejected only 0.3\% of the time (as expected for a 99.7\% significance threshold). The solid contours correspond to the number density of pulsars (in units of pulsars per square degree) as computed using Eq. \ref{eqn:spdef}. 

There are two competing factors which shape the transition between the sensitive and insensitive regions of parameter space. The plateau at small values of $\g$ represents the limit of low numbers of pulsars. Obviously, if there are no pulsars in the sky there is no signal to be detected. Within the flux range explored here the $A$ test is not sensitive if there are fewer than  $\sim 10$ pulsars in the 40,000 pixels.

The vertical transition is explained by the fact that pulsars must contribute the {\em highest} peak in the power spectrum in order to be detected by the periodicity test. As the flux of each pulsar is increased (moving to the right in Fig.~\ref{fig:40k}) the power spectrum peak at the pulsar's frequency will eventually become the highest peak in the power spectrum. This then causes the non-Gumbel-ness of the pixel scores which is detected by the $A$ test.

We can view this as a requirement that the quantity $P_p$ (Eq. \ref{eqn:Pp}) be comparable to $\log\Nbins$, the mode of the distribution for the maximum normalized power in the case of no pulsars. The $\alpha \, \SN^2$ term in Eq. \ref{eqn:Pp} is most important in governing the transition. Because the Gumbel distribtion only contains a location parameter we can write an approximate equation describing the vertical part of the sensitivity transition:
\begin{equation}
\alpha \, \SN^2 \simeq \log\Nbins.
\label{eqn:trans}
\end{equation}
The left hand side is an estimate of the height of the peak corresponding to the actual pulsar signal. The right hand side is the maximum power in the other $\Nbins -1$ frequency bins. Only when the left hand side is greater than the right hand side will the method be able to reject the null hypothesis of no pulsars. This is because the periodicity statistic we have chosen is not sensitive to pulsar peaks which are subdominant in the power spectrum.

The photon fluxes of individual pulsars in the simulated parameter space are all below the point source sensitivity of the LAT \citep{2009ApJ...697.1071A}. The pulsars in the simulation would be undetected by Fermi as bright sources. Therefore, ``blind searches'' would not consider these pulsars as candidates for periodicity searches. Such objects truly contribute to the diffuse background.

Nevertheless, if we measure the power spectrum for a pixel which contains an unresolved pulsar the spike at the pulsar's frequency might be large enough to constitute a detection of a periodic source. To estimate when this occurs we consider the following hypothesis test.  We measure a power spectrum peak height of $x$, and ask ``What is the probability that at least one peak in any of the observed time series has a value of $x$ or higher {\em if} there were no periodic sources present in the data?'' The answer again follows from the Gumbel distribution (Eq.~\ref{eqn:CDF}) but with $\Nbins$ replaced with $\Nbins \times \Npix$, i.e. the number of independent frequency bins for each time series multiplied by $\Npix$, the number of time series considered (in this case 40,000). The quantity $F(x)$ is the significance of this peak.

In the region to the right of the dashed line in Fig.~\ref{fig:40k} individual pulsars would be detected at $5\sigma$ based on the height of the power spectrum peak derived from their pixel's time series. The region's shape is governed by an equation similar to Eq.~\ref{eqn:trans} except that the right-hand side is replaced by a peak height $x_{5\sigma}$ such that $1-F(x_{5\sigma}) \simeq 5.7 \times 10^{-7}$, corresponding to a $5\sigma$ detection. This suggests that {\it simply computing the power spectra for the entire sky may turn up detections of pulsars which are too faint to be flux-resolved}.

\begin{figure}
\begin{center}
\includegraphics[height=7.cm]{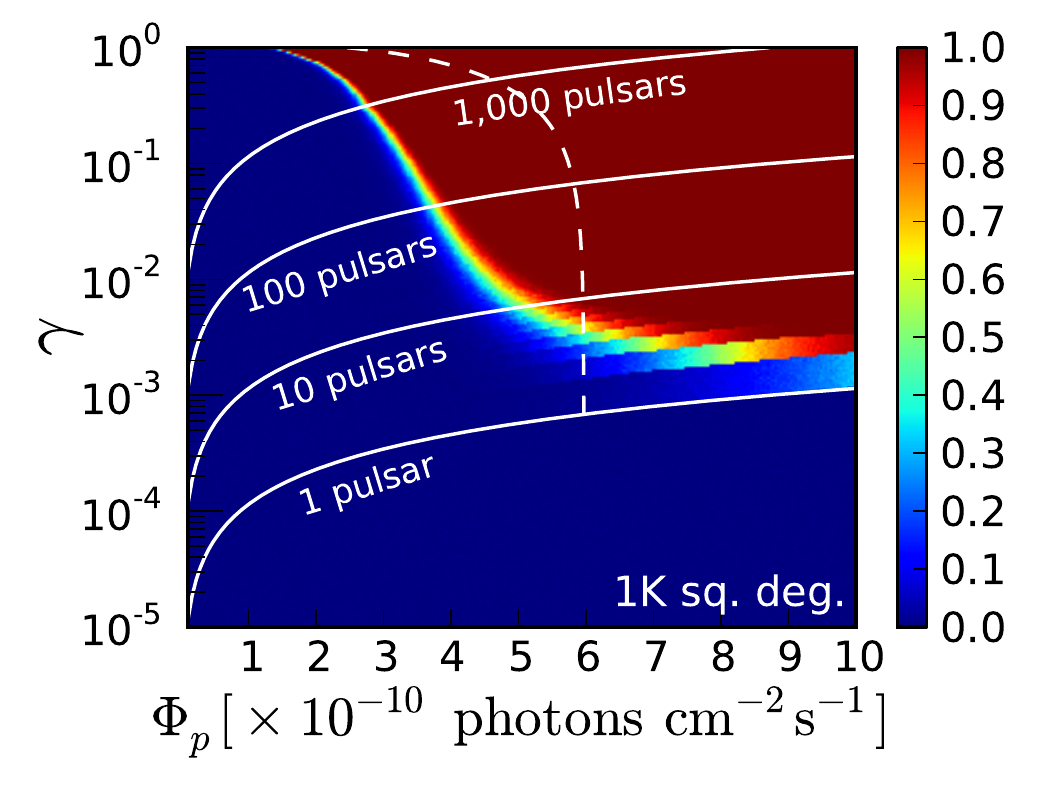}
\caption{Same as Fig.~\ref{fig:40k} but for an observation area of 1,000 square degrees corresponding to a study of the Galactic center. Here, solid contours depict the total number of pulsars present in the observed region. The dashed line denotes the $5\sigma$ detection threshold of individual pulsars based on power spectrum peak height as in Fig.~\ref{fig:40k}.}
\label{fig:1k}
\end{center}
\end{figure}

In Fig.~\ref{fig:1k} we show the results of a similar simulation but for an observation of only 1,000 square degrees of sky. This situation represents a study of the galactic center, a region whose source population is of great interest \citep{2011arXiv1106.3583B, 2011PhLB..697..412H,2011JCAP...03..010A,2011arXiv1102.5095D,2010arXiv1012.5839B,2011PhRvD..83h3517H}. The shape of the sensitivity region is similar to the all-sky survey. The stripe pattern is caused by the discrete addition of pulsars to the survey area. {\it The method is sensitive to the presence of pulsars even when individual pulsars are unresolved in both flux and Fourier power}.
The sensitive (dark red) region is larger in the all-sky survey than in the galactic center study. This demonstrates that the statistical test benefits from larger numbers of measured time series (assuming equal fluxes and number densities of pulsars).

\section{Discussion}
\label{sec:discussion}


Using the approximation to the sensitivity transition given by Eq.~\ref{eqn:trans} we can estimate how the result of the simulations discussed in the previous section will scale with changing parameters. The Gumbel distribution we have been exploring has the beneficial property that the location parameter goes as the logarithm of the number of independent trials. We expect this to be a general feature of any periodicity test. Thus, as observation time $T$ increases the right hand side of Eq. \ref{eqn:trans} increases as $\log T$. The left-hand side increases in proportion to $\Fp^2 \Aeff T / \O$. Thus, this technique benefits from longer observation times, larger effective areas, and smaller pixel sizes (i.e. future gamma-ray observatories) in the same ``root N'' way that conventional searches do.


The main difficulty in the outlined strategy lies in choosing a good test of periodicity and in the computational challenge of computing it many times for the different time series. Traditionally, pulsars are searched for either by taking a Fourier transform of the time series or by folding the time series in the time domain at many different trial periods. 

Regular pulsars do not have constant periodicity but experience spin-down (magnetic braking) and glitches. These complicating factors force statistical tests for periodicity to be performed on a large grid of trial frequency derivatives or on short stretches of the time series \citep{2006ApJ...652L..49A,1987A&A...175..353B}. In Fourier space the changing period of the pulsar acts to spread its signal power over many frequency bins, diluting the peak amplitude. Millisecond pulsars, on the other hand, have extremely stable rotations, with period derivatives on the order of $10^{-19}$ s/s \citep{2008LRR....11....8L}. Even over observation periods of years the frequency of many MSPs will not drift into neighboring Fourier bins \citep{2007AIPC..921...54R}.  Thus, the number of trials performed when computing the test statistic can be significantly lower than for regular pulsars.

Additionally, MSPs are thought to form in binary star systems. Because binary systems are more common than single stars most galactic pulsars are likely members of a binary pair that have been spun up into MSPs \citep{1982Natur.300..728A,1982CSci...51.1096R,1991PhR...203....1B,2006csxs.book..623T}   In addition, it has recently been suggested that MSPs might dominate normal pulsars in their contribution to the gamma-ray background \citep{2010JCAP...01..005F}. Millisecond pulsars are also older, have had more orbital trips around the galaxy, and therefore are more likely to be found at higher galactic latitudes than normal pulsars. Therefore, these pulsars may be important contributors to the so-called ``extra-galactic'' or isotropic gamma-ray background \citep{2010arXiv1011.5501S}.

\subsection{Caveats and Improvements}

We have been optimistic in some areas and overly simplistic in others. Here we review some of the practical difficulties in performing this test on LAT data and point out the simplifications we have made and how they affect the results. 

The most obvious difficulty we have glossed over is the fact that pulsar light curves are not simple sine waves. This has the effect of dispersing the power in the frequency bin centered at the pulsar's frequency into higher harmonics. The simple periodicity test we proposed (maximum normalized power) is almost certainly not optimal for the case when power is found at higher harmonic frequencies (see below for ways to try to recover this power). We have left room in the analysis (see Eq.~\ref{eqn:Pp}) for a reduction of $\alpha$, designed to account for the effect of power being dispersed into other frequency bins. Equation \ref{eqn:trans} suggests that the pulsar flux one is sensitive to goes as $1/\sqrt{\alpha}$ (since $\SN$ scales proportionally to $\Fp$).

While millisecond pulsars are extremely stable and do not experience glitches or suffer from rapid spin-down, their rotation is not completely constant. It is therefore probable that some power is dispersed into neighboring frequency bins by non-negligible period derivatives. The techniques used to try to recover this power involve performing many analyses with different trial period derivatives. Specifically, the arrival times of the photons are corrected to account for a spin-down effect and then the periodicity search is performed on this modified time series. The decrease in sensitivity due to spin-down increases as the observation time increases.

As millisecond pulsars are found in binary systems, the orbital motion of the binary can cause distortions in the Fourier spectrum of the time series. Essentially, the orbit of the pulsar causes a doppler shift in its period which disperses Fourier power into different frequency bins. Methods have been proposed that can sweep up this power~\citep{2003ApJ...589..911R}. Such methods can be incorporated into a more advanced periodicity test.

Errors in source position are known to affect the detectability of individual pulsars. The first step in analyzing a time series is to correct for motion of the detector with respect to the pulsar. This correction depends on an accurate ``barycentering'' procedure, which in turn relies on precise knowledge of the pulsar's position. In searching the background for unresolved pulsars we have no information as to  where the pulsars are located within the pixels, and this affects the quality of the barycentering.

The most important consideration that goes into a realistic application of the proposed method is the choice of periodicity statistic. In practice, one is bound by finite computational resources --- ideally, one would perform a detailed time series analysis on every pixel in the sky, including searching over trial periods, period derivatives, and other ephemera.
We have been simplistic in the choice of the maximum normalized power as a test statistic. A first generalization is to search harmonic sums of the normalized power spectrum. This would take into account that pulsar light curves are not sine waves. Considering the harmonic sum of the power spectrum is an attempt to recover the as much signal power as possible. The statistics of such a test are relatively straightforward to compute.

In addition, there are several choices of tests for periodicity currently in use to search for pulsars in radio data and in gamma-rays. The H test \citep{1989A&A...221..180D} and the $\z22$ test \citep{1983A&A...128..245B} are based on binning the photon arrival times by phase for a given trial pulsar period and then checking whether the distribution of phases is consistent with random. These tests require a guess for the pulsar period. However, it is computationally intensive to calculate the statistic for every possible value of the pulsar period for a large sample of pixels. More recently, a time-differencing technique \citep{2006ApJ...652L..49A} has been proposed to overcome some of these computational challenges and has been very successful in discovering new pulsars with Fermi-LAT \citep{2009Sci...325..840A, 2010ApJS..187..460A}.

To adapt these tests to the present task, we propose to first find the power spectrum of the time series (or its harmonic sum) and take the $n$ highest peaks to be trial periods for the more advanced algorithms. The number of trials $n$ would need to be adjusted based on computational resources and the choice could be calibrated by examining the power spectra from time series which are known to contain gamma-ray pulsars. An automated analysis pipeline can be conceived in which one would perform a cursory scan of the time series looking for semi-significant peaks and then perform additional, computationally intensive scans of these peaks, assigning a periodicity score at the end.

Besides computational cost one has to balance two factors when deciding on a periodicity test. The test should be as sensitive as possible to the presence of periodic signals but should also minimize the number of ``trials''. A large number of trials raises the possibility that a random signal, by chance, could appear periodic. In our case the number of trials was the number of independent Fourier bins that were scanned when looking for peaks. As the number of ``trials'' grows it is more likely to find a random outlier that mimics periodicity.

Any periodicity test or analysis procedure can be adapted to the search for unresolved pulsars. The key ingredient is the null distribution of the periodicity scores. For example, an arbitrarily complex analysis pipeline can be established which takes a time series and outputs a periodicity score. The inner-workings of the pipeline can involve scanning over trial periods and period derivatives. It can include identifying promising peaks for more careful scanning. Once the procedure is set, one simply runs it many times on uncorrelated photon time series (i.e. white noise). The resulting set of periodicity scores constitutes the null distribution. The pipeline is then applied to actual measured time series and the resulting scores are collectively checked for inconsistency with the null distribution.

We can illustrate the effect of different tests using the sensitivity plot in Fig.~\ref{fig:40k}. Different periodicity tests would shift both the sensitive (red) region and dashed line together. However, the scaling is not necessarily fixed. The dark red region of parameter space to the left of the dashed line remains the most interesting. It is here where searches for individual pulsars would fail but where the collective statistics would succeed in revealing their presence. The size and shape of this region likely depends on which periodicity test is chosen. We plan to explore other tests in future work.

Furthermore, the division of a region of the sky into spatially separated time series (step one in Section 2.1) can also be optimized. Instead of breaking the sky into pixels and taking the time series of each one, an alternate technique is to only search promising sky locations for evidence of periodicity. One could consider only ``bright spots'' or ``hot pixels'', regions of the sky with a signal to noise ratio greater than 1, say. Alternatively, the candidate locations can be chosen from lists of known sources (see Fermi bright source list, \cite{2009ApJS..183...46A}, \cite{2010ApJS..188..405A}), or from pulsar candidate locations in blind searches. The later have been previously analyzed for pulsations but have not been {\em jointly} searched for unresolved pulsars. These strategies have several advantages. The computational burden would be reduced because of the fewer number of time series to scan. The barycenter correction would be improved by the better localization of the sources' positions. A priori, {\it hot pixels have a higher chance of containing pulsars than randomly selected pixels, leading to a larger fraction of the searched pixels that contain pulsars} (effectively increasing $\sp$ in Fig.~\ref{fig:40k}).

Because the analysis is sensitive only to the highest power spectrum peak it is almost completely insensitive to the possibility that there may be multiple pulsars contributing to a single time series. However, this situation likely occurs in globular clusters and in the galactic center region, both places conceivably containing important populations of pulsars. A periodicity statistic should be tailored specifically to studies of these regions. A simple generalization of the periodicity test would be to take the top $n$ highest peaks in each time series instead of just the highest. Then we would have $n$ periodicity scores from each pixel instead of one. Alternatively, one could count the number of peaks with height greater than some threshold. The score from each pixel would be this integer number. (In both cases the search could take place using the harmonically summed power spectra.)

\subsection{Pulsar population parameter estimation}

This analysis begs the followup question of how we can learn the details of the pulsar population from studies like this, where individual pulsars remain undiscovered. In particular, it is of great interest to determine what fraction of the gamma-ray background is due to unresolved pulsars (the value of the quantity $\g$ in the simulations of Sec.~\ref{sec:fermi}). The detailed extraction of population parameters from the collection of periodicity scores requires some kind of modeling of the population. However, we can use the simplified model presented here to place interesting constraints on the number of pulsars with certain fluxes without any detailed modeling.

In the simulations of Sec.~\ref{sec:fermi} we assumed that every pulsar had the same flux. This is obviously false if we claim that the simulated pulsars make up all the pulsars in the sky. However, the simulated pulsars can instead be interpreted as a ``slice'' of the number function of pulsars.

An important description of the pulsar population is the number density of pulsars $\sp(\Fp)$ with flux greater than $\Fp$. This function can be used to define the total contribution from pulsars:
\begin{equation}
\g = \frac{1}{\Ftot} \int\limits \Fp \left|\frac{d \sp}{d \Fp}\right|d \Fp.
\label{eqn:NF}
\end{equation}

The simple simulations of Sec.~\ref{sec:fermi} can be used to constrain $\sp(\Fp)$ as follows. Imagine that we have performed a test over the whole sky (Fig.~\ref{fig:40k}) but failed to reject the null hypothesis. At each flux $\Fp$ we can draw a line straight upwards in Fig.~\ref{fig:40k} until we reach the transition to the dark red region. Let the number density of pulsars simulated at this transition point be given by $\tilde{\sigma}_p(\Fp)$. Then we can claim that the true number density function at this flux $ \sp(\Fp)$ must be less than $ \tilde{\sigma}_p(\Fp)$. If this were not the case then there is a 99.7\% (in this example) chance that the statistical test would have detected the presence of these pulsars. This constraint relies on the choice of $\alpha$ and in practice the choice should be calibrated using known pulsar light curves.

If we are willing to make some assumptions about the shape of $ \sp( \Fp)$ and only allow its overall normalization to vary we can make stronger statements. In this case we could actually simulate a population of pulsars for different choices of normalization and find the sensitivity of the method to each choice. The test will become sensitive above some critical value of the normalization. Depending on whether the test rejects or does not reject the null hypothesis we could then place a lower or upper bound on the normalization of the number density function. This bound would then immediately translate into a bound on the total contribution of pulsars to the background (Eq.~\ref{eqn:NF}). There are several motivated choices for the shape of $ \sp(\Fp)$ which depend on the spatial distribution of pulsars \cite{2010JCAP...01..005F}. In reality, however, the population of gamma-ray pulsars is completely unconstrained at fluxes below about $10^{-8}$ photons ${\rm cm}^{-2}{\rm s}^{-1}$ \citep{2010ApJS..187..460A}.

In addition, one can analyze the measured distribution of periodicity scores using conventional $\chi^2$ minimization. In this case it is necessary to know what the distribution of scores will be as a function of the pulsar population parameters. One then can bin the measured scores and find the best fitting population parameters. The pulsar population models can be made as complicated as one likes --- the analysis requires a scan over this parameter space looking for regions whose score distribution matches the observed one. We defer applications of these techniques to the LAT data in future work.

\section{Conclusions}

In this manuscript we propose a new technique whose application to Fermi-LAT data can reveal the extent to which pulsars contribute to the gamma-ray background. The method is based on the cumulative statistics of photon time series that are binned spatially. The motivation behind this approach lies in the general idea that even though individual pulsar searches may be unsuccessful, information from undetected pulsars is still measurably encoded in the gamma-ray background.

In general, current pulsar searches are based on the evidence of a source at a particular location. These sources are subjected to a battery of periodicity tests, and careful analysis of LAT data has already revealed the presence of gamma-ray pulsars. However, it is likely that large numbers of pulsars are beyond the current reach of LAT to even identify their associated events. These pulsars (with very weak signals) will contribute to the diffuse gamma-ray background. 

Our main results are:
\begin{itemize}
\item The proposed technique has the ability to discover a pulsar contribution to the gamma-ray background if the fraction due to pulsars is greater than $10^{-3}$.
\item It is sensitive to a population of pulsars whose individual photon fluxes are as low as $10^{-10}$ cm$^{-2}$ s$^{-1}$.
\item Using the photon time series derived from a specific location on the sky, one can discover {\it individual} pulsars with photon fluxes down to about $6\times 10^{-10}$ cm$^{-2}$ s$^{-1}$, which is below the current point source sensitivity threshold.
\item By considering only ``hot pixels'' or current blind search candidates the sensitivity of the method is increased markedly. 
\item Any periodicity test or analysis pipeline can be applied to the search for the unresolved population. 
The only requirement is the response of the test to  uncorrelated photon time series. This allows the technique to be optimized for any  given application (e.g. all-sky surveys, galactic center, globular clusters, etc.).
\end{itemize}

The method proposed in this work takes advantage of all events in the diffuse gamma-ray background and gives information about the population of unresolved pulsars. The importance of this task goes beyond pulsar astrophysics. It is manifestly apparent that a detailed understanding of astrophysical backgrounds is vital in any gamma-ray observation, including surveys of astrophysical sources (e.g., blazars), as well as studies of more exotic and hypothetical contributions (e.g., annihilating dark matter). It is therefore of extreme interest to apply this technique to current and future gamma-ray data.

We acknowledge useful conversations with Jacqueline Chen, Ian Dell'Antonio, Andrew Favaloro, Scott Field, Richard Gaitskell, Elizabeth Hayes, Kyle Helson, Julie McEnery, Igor Moskalenko, Troy Porter,  Bob Sameth, and Jason Su. SMK thanks the Aspen Center of Physics for hospitality during the early stages of this work. SMK and AGS are funded by NSF PHYS-0969853 and by Brown University.

\bibliographystyle{mn2e}

\bibliography{manuscript}

\appendix
\section{The $A$ test}

In this appendix we provide details about the statistical test we used in this paper. The test is designed to determine if a collection of observations is inconsistent with having been drawn from a given null distribution. It is meant to be sensitive to a small upper tail in excess of what is predicted by the null distribution. Although motivated by the application to pulsars the $A$ test has nothing to do with astrophysics and may be used in any statistical study.

\subsection{Motivation}

Recall the situation presented in the text. We have a collection of periodicity scores (denoted ${x_i}$) and want to test whether the collection is consistent with having been drawn from the null distribution (in this case a Gumbel distribution). The goal is to boil the collection of scores down into a single number $A$ and then study the distribution of $A$ under the null hypothesis. The quantity $A$ is meant to reflect the overall level of periodicity in the sample.

The critical value $A^*$ is defined by the property that, if the null hypothesis is true, the probability that $A$ is less than $A^*$ is e.g., 99.7\%. To be precise, $A$ is a function of the collection ${x_i}$. If the $x$'s are each drawn from the null distribution then the probability that $A$ is less than $A^*$ is 0.997, or whatever the desired significance is.

Different choices of $A$ may be more or less powerful. In general, the power of a test is defined as the probability that the null hypothesis is rejected when the null hypothesis is, in fact, false. If it is unlikely that $A$ is above some critical value $A^*$ even when there are many pulsars present in the sky a poor definition for $A$ has been chosen. Unfortunately, only in special, simple cases is there a ``uniformly most powerful'' test. In the particular case we are studying here there are many degrees of freedom associated with the alternative hypothesis. For example, the light curves of pulsars and their number density are functions which must be specified by many parameters. As a result there is no uniformly most powerful test in this case. (See e.g.~\cite{Kendall5th} for an more detailed discussion.)

In order to choose a powerful statistical test we must examine the behavior of the collection of $x$'s in the case where pulsars are present. Consider a pixel which contains a pulsar. The only way the $x$-value of this pixel will contain any information about the pulsar is if the peak in the normalized power spectrum is actually due to the pulsar. Under the null hypothesis, each $x$ is drawn from the Gumbel distribution in Eq. \ref{eqn:CDF}. The effect of pulsars is to skew the distribution towards higher values of $x$: the pixels with a pulsar have a chance of replacing the peak power in a random frequency bin with the power at the pulsar's frequency. Based on these considerations we would like to choose a statistical test that puts more weight on higher $x$ values. 

There are a wide variety of statistical tests that are in common use. The Kolmogorov-Smirnov (KS) statistic is commonly used in astronomy. Kuiper's extension of the KS statistic gives more weight to the tails of the distribution. This would be beneficial for looking for the excess in large $x$-values. The Anderson-Darling statistic is used more rarely but also gives extra weight to the tails. Likelihood ratio statistics are another option, though these require some knowledge of the alternative hypothesis that one is testing for. It is known that likelihood ratio tests are the most powerful tests for ``point'' hypotheses \cite{Kendall5th}. They are based on the likelihood function for the data under various hypotheses, and should therefore exploit all the information available in the data.

The proposed $A$ test statistic is designed to be sensitive to the upper tail of a distribution. It shares properties with the Anderson-Darling and KS tests and can also be interpreted as a likelihood-ratio test. Unlike these other tests, however, the distribution of the $A$ test statistic under the null hypothesis is very simple (a gamma distribution). It is expected to be powerful (like a likelihood test) but also very easy to use (no sorting of the data and no lookup tables).

\subsection{Details}

In this subsection we present the details of the $A$ test. The task is to take a collection of numerical values and determine if this collection is consistent with being drawn from a given probability distribution (the null distribution). Below, this collection of numbers will also be referred to as the ``data'' or the ``samples''.

When looking for an extended tail in a collection of measured quantities we noticed that it is often useful to look at the logarithm of the empirical survival function (SF) of the data. The empirical SF is defined as $1-F_N(x)$, where $F_N(x)$ is the empirical cumulative distribution function (CDF). Simply put, the SF at some value $x$ is the fraction of the sample values which are greater than $x$. Thus, at $x=-\infty$ the empirical SF equals 1 and decreases by $1/N$ every time $x$ crosses one of the measured values, where $N$ is the sample size. This empirical SF can be compared to the theoretical SF for the case where the data come from the null distribution. For the null distribution, the survival function is simply $1-F(x)$, where $F(x)$ is the usual cumulative distribution function for the null distribution.

When comparing the logarithm of the empirical and theoretical SFs any excess at large values of $x$ becomes more pronounced, even if only a small fraction of the samples are at such large values. Therefore, we order the data by increasing $x$-value and define the $A$ statistic as
\begin{equation}
A \equiv \frac{1}{\sqrt{N}} \sum\limits_{i=1}^{N} \left\{ \log\left[1-F_N(x_i)\right] - \log\left[1-F(x_i)\right] \right\},
\label{eqn:Adef1}
\end{equation}
where $x_1 < x_2 < \dots < x_N$ and $F_N(x_i) = (i-1)/N$, the empirical CDF. We can make some simplifications to the first term in the sum as :
\begin{eqnarray}
\sum\limits_{i=1}^{N} \log\left[ 1-F_N(x_i)\right] &=& \sum\limits_{i=1}^{N} \log\left[1-\frac{(i-1)}{N} \right] \nonumber \\
&=& \log\left[ 1 \left(1-\frac{1}{N}\right) \left(1-\frac{2}{N}\right) \cdots \left(1-\frac{N-1}{N}\right) \right] \nonumber \\
&=& \log\left[ N \left(N-1\right)\left(N-2\right)\cdots 1 \right]- \log\left[N^N\right] \nonumber \\
&=& \log N! - N\log N.
\label{eqn:Aconst}
\end{eqnarray}
Inserting this back into the definition of $A$ we have
\begin{equation}
A = \frac{1}{\sqrt{N}} \left\{ \left[ \sum\limits_{i=1}^{N} -\log\left[1-F(x_i) \right] \right] - N\log N + \log N! \right\},
\label{eqn:Adef}
\end{equation}
The statistics of $A$ is governed by the term in curly brackets. In this sum the numerical ordering of the $x$'s does not matter since the sum is over all of them. The distribution of $A$ under the null hypothesis is now straightforward to find. For any random variable $X$ with CDF $F$ the quantity $F(X)$ is distributed uniformly in the interval between 0 and 1. This implies that $1-F(X)$ is also uniformly distributed on this interval. Now, the negative logarithm of such a uniformly distributed variable is distributed according to the exponential distribution with scale factor 1. Therefore, under the null hypothesis the quantity
\begin{equation}
G \equiv \sum_{i=1}^{N} -\log[1-F(x_i)]
\label{eqn:Gdef}
\end{equation} 
is the sum of $N$ exponentially distributed random variates. This sum is described by the well-known gamma distribution (also called the Erlang distribution in this case) with shape parameter $N$. The inverse CDF of the gamma function then provides the critical value $A^*$. For instance, to find the value of $A^*$ under which there is a 99.7\% chance of measuring $A$ (under the null hypothesis) one determines the value of $G^*$ that satisfies
\begin{equation}
0.997 = \int_{0}^{G^*} \frac{x^{N-1} e^{-x}}{(N-1)!} \, dx.
\end{equation}
The quantity $G^*$ is then inserted into Eq.~\ref{eqn:Adef}, replacing the term in curly brackets. The resulting value of $A$ is $A^*$. If for a given sample of $N$ $x$-values the quantity $A$ (Eq. \ref{eqn:Adef}) is greater than $A^*$ then one can reject that the sample came from the distribution with CDF $F(x)$ at 99.7\% significance.

\subsection{Properties of $A$}

Of course, there is no reason to include the constant terms in Eq. \ref{eqn:Adef}. One can just take the test statistic to be $G$ (Eq.~\ref{eqn:Gdef}), the only quantity that depends on the data. Then $G^*$, discussed above, is the critical value for the test statistic. (In fact, this is how we actually performed the simulations.) However, the definition we have given for $A$ (Eq.~\ref{eqn:Adef}) has a nice asymptotic property for large sample sizes (i.e. as $N \rightarrow \infty$). The central limit theorem says that the gamma distribution converges to a normal distribution with mean $N$ and standard deviation $\sqrt{N}$. In the same limit the constant term Eq. \ref{eqn:Aconst} converges to $-N$ as can be seen using the approximation for $\log(N!)$ found in every statistical mechanics textbook (e.g. \cite{1965fstp.book.....R}, section A.6). Therefore as $N \rightarrow \infty$ the distribution for $A$ converges to a standard normal distribution (i.e. normal with mean 0 and variance 1).

The $A$ test statistic is similar to the KS and Anderson-Darling statistics in that is based on the CDF of the null distribution. The CDF has the nice property that it is distributed uniformly (if the null hypothesis is true). This allows the null distributions for the KS, Anderson-Darling, and $A$ test statistics to be found analytically.

The specific application of the $A$ test statistic shown in this paper can also be interpreted as a likelihood-ratio test. The null distribution is given by the Gumbel distribution with a peak at $\log \Nbins$. Imagine that the alternative distribution for the $x$'s follows the null distribution for values of $x$ less than $\log \Nbins$ but does not fall off for higher values. This is supposed to represent the situation when pulsars are present: there are more large values of $x$. The likelihood ratio is the ratio of the alternative PDF to the null PDF (as functions of $x$). When this quantity is large it indicates that the alternative describes the sample better than the null does. The likelihood ratio is the product of these ratios for each $x_i$. It is usually easier to work with the logarithm of this quantity which is the sum of the logarithms of the individual likelihood ratio terms.

Let us see how each term in the log-likelihood ratio compares to each term in the $G$ statistic (i.e. each term in the curly bracketed sum in Eq. \ref{eqn:Adef}). If $x$ is less than $\log\Nbins$ both statistics contribute approximately 0. In the case of the likelihood ratio this is because the null and alternative PDFs are defined to be the same there (so the $\log$ of their ratio is $0$). It is also easy to see from Eq. \ref{eqn:CDF} that when $x$ is less than $\log\Nbins$, $F(x)$ is close to 0. If $x$ is greater than $\log\Nbins$ the quantity $1-F(x)$ becomes approximately $\exp(-(x-\log\Nbins))$ and so $-\log(1-F(x)) \simeq x-\log\Nbins$. For the likelihood ratio when $x > \log\Nbins$ the alternative hypothesis PDF is 1 and the null PDF is approximately $\exp(-(x-\log\Nbins))$. Thus the logarithm of this ratio is also approximately $x-\log\Nbins$.

For all values of $x$, therefore, the $A$ statistic (based on the quantity $G$) behaves just like a likelihood ratio test that is designed to pick up an extended upper tail in the sample. This implies that the $A$ test should be a powerful test in looking for such a tail. Moreover, the null distribution of $A$ has a particularly simple form (a shifted and scaled gamma distribution) and converges to the standard normal distribution when the sample size is large, making $A$ an attractive addition to the current library of tests.

\end{document}